\documentclass[
reprint,
superscriptaddress,
showkeys,
aps,
prd
]{revtex4}

\usepackage{babel,comment}
\usepackage{diagbox}
\usepackage{soul}
\usepackage{amsmath,amsfonts,mathrsfs,amssymb}
\usepackage{epsfig}
\usepackage{latexsym}
\usepackage{graphicx}
\usepackage{dcolumn}
\usepackage{bm}
\usepackage{bbm}
\usepackage[T1]{fontenc} 
\usepackage{epsf}
\usepackage{amsthm}
\usepackage{color}
\usepackage{slashed}
\usepackage{float}
\usepackage{esvect}
\usepackage{mathtools}
\usepackage{hyphenat}
\usepackage[pdfencoding=auto]{hyperref}
\newcommand{\RNum}[1]{\uppercase\expandafter{\romannumeral #1\relax}}

\usepackage{cancel,soul,ulem}
\renewcommand{\sout}{\bgroup \color{red} \ULdepth=-.5ex \ULset}
\begin{document}

\title{Temperature derivative divergence of the electric conductivity and thermal photon emission rate at the critical end point from holography}

\author{Yi-Ping Si}
\email[]{siyiping23@mails.ucas.ac.cn}
\affiliation{School of Physical Science, University of Chinese Academy of Sciences, Beijing 100049, China}

\author{Danning Li}
\email[]{lidanning@jnu.edu.cn}
\affiliation{Department of Physics and Siyuan Laboratory, Jinan University, Guangzhou 510632, China}

\author{Mei Huang}
\email[]{huangmei@ucas.ac.cn}
\affiliation{School of Nuclear Science and Technology, University of Chinese Academy of Sciences, Beijing 100049, China}

\begin{abstract}
The thermal photon emission rate $\frac{d\Gamma}{dk}$ and DC eletric conductivity $\sigma_{Q}$ of the strongly coupled quark-gluon plasma (sQGP) are investigated around the critical end point in a $N_f=2+1$ holographic QCD model with parameters obtained from machine-learning. It is found that both thermal photon emission rate and eletric conductivity grow most obviously around $T_c$, which agrees with the previous studies, and the result of eletric conductivity at zero chemical potential resembles the lattice results. Moreover, it is found that both the temperature derivative of the  eletric conductivity and thermal photon emission rate diverge at the critical end point.  
\end{abstract}

\maketitle

\section{Introduction}

The existence and location of the critical endpoint (CEP) in the QCD phase diagram has been widely investigated theoretically \cite{Pisarski:1983ms,Asakawa:1989bq,Stephanov:1998dy,Hatta:2002sj,Stephanov:1999zu,Hatta:2003wn,Schwarz:1999dj,Zhuang:2000ub}. The search for CEP serves as a prominent goal of heavy ion collision (HIC) experiments in facilities like RHIC at BNL, FAIR at GSI, NICA at DUBNA, and HIAF in Huizhou. The non-monotonic behavior of baryon number fluctuations has been considered as a signature of the CEP \cite{STAR:2010-CEP,Exp-CEPsearch-2017,STAR-2020-Non-monotonic,STAR2022-cumulants}. The latest STAR results
\cite{Ashish2024,STAR:2025zdq} didn't show clear evidence of the existence of CEP for collision energy $\sqrt{s}>$ 7.7 GeV. From the analysis in Ref.\cite{Shen:2024ptz}, the baryon number fluctuation signature might be distorted by the hadronization process. Therefore, a different signature of the CEP which is insensitive to the hadronization would be needed.

For the quark-gluon plasma (QGP) produced in the HICs, one important class of observables is the electromagnetic probes, which mainly include photons and dileptons. The thermal-photon and dilepton produced in heavy-ion collisions are important observables in studying the properties of QGP \cite{David2019-Photon-Dilepton-rev,Geurts2022-Photon-Dilepton-rev02}. Due to their colorless nature and relatively long mean free path, once generated, they pass through the fireball with barely further interaction. Thus the information about the early stage of the collision and its subsequent evolution is well preserved by those electromagnetic probes. Photons produced in HIC can be divided into different types, including prompt (produced in early hard patron scattering), thermal (produced in QGP and latter hadron phases) and decay (from the deacy of final-state hadrons) photons. The prompt and thermal photons are of particular interest experimentally, and are referred to as "direct" photons. However, photons produced in various stages are detected collectively. Therefore, it's necessary to differentiate photon production from different sources. Thermal photons mainly contribute to the total production in the low transverse momentum region. The observed excess of the direct photon in Au+Au (PHENIX) \cite{PHENIX:2008uif} and Pb+Pb (ALICE) \cite{Wilde:2012wc} is attributed to thermal photons. 
Much theoretical work has been devoted to calculating the production of direct photons, including perturbative QCD (pQCD) calculations for leading order (LO) \cite{pQCD-LO2001} and next-to-leading order (NLO) \cite{pQCD-NLO2013}, lattice simulations \cite{latticePhoton-2022,HotQCDCollaboration:2024-latticePhoton}, and results from effective models such as holographic models \cite{Caron-Huot,Stefano2015,YangDLConductPLB}. However, a clear understanding of the possible signature of the CEP from photon emissions has not yet been concluded.

Apart from photon emission, the transport properties are necessary for the understanding of QGP and the behavior near critical endpoint as well. Many effective models are constructed to study the real time evolution of the plasma (e.g. Boltzmann transport, hydrodynamics and hadronic cascade \cite{Molnar:2001-Boltzmann-Transport,Gale:2013-Hydro-Simulation-Review,Bleicher:1999-parton-cascade-01,Bass:1998-parton-cascade-02}); the transport coefficients serve as parameters in those models. For example, it has been shown that strong electromagnetic fields are generated in the HIC process, and the electric conductivity, the transport coefficient which descibes how the medium reacts to an external electric field, is crucial for such simulations. Except for phenomenological applications, the transport coefficients themselves characterize the nonequilibrium dynamics of strongly coupled plasmas and the behaviors of which can also indicate phase transition. It has been shown that an enhancement of bulk viscosity occurs around $T_c$ \cite{Kharzeev:2007-bulk-viscosity-at-TC-0mu-lat,Karsch:2007-bulk-viscosity-at-TC-0mu-lat,Karsch:2007-bulk-viscosity-at-TC-03} and the claim that shear viscosity can indicate phase transition has been made as well \cite{Lacey:2006bc-shearAsCEPsign}. Similarly, the electric conductivity of QGP which is directly related to the photon and soft dilepton production, also has the possibility to serve as a signal of phase transition. In fact, it has already been shown by both lattice results \cite{GertAartsAmato2013PRL,GertAartsAmato2014JHEP,2019ELEconductivityLattice} (for a review see \cite{GertAarts2020}) and various phenomenological studies \cite{Stefano2013,Stefano2015,I.I.YangKiritisis:2016ugz,ArefevaRussia2021} that a "rapid growth" or "jump" occurs at the transition temperature $T_c$. For the crossover region, it's pseudo-transition temperatures, but for simplicity we'll still call them $T_c$ here and below. Moreover, a series of recent work based on NJL model \cite{Nishimura:2023-diverge-conductivity,Nishimura:2024-diverge-conductivity} predicts that the electric conductivity will diverge at the CEP. Besides, in spite of the abundant studies, there's no consensus on the magnitude of the electric conductivity yet \cite{Huang:2022qdn-uncertainElectricConduct}. This suggests that the research on the electric conductivity is far from settled.

Although much work has been done to investigate the photon production through pQCD and lattice calculations, as for the search for CEP, these results are very limited in the region of interest. Specifically speaking, QGP around CEP is strongly coupled and thus the perturbative methods are not applicable, while lattice calculation is limited within zero or small baryonic chemical potential regions due to the "fermion sign" problem. On the other hand, the gauge/gravity duality (GGD) \cite{Maldacena1997-GGD-01,Witten1998-GGD-02,Gubser1998-GGD-03} has emerged as a powerful tool to investigate the strongly coupled QGP. The duality claims that a strongly coupled gauge theory in four-dimensional spacetime is dynamically equivalent to a gravitational theory in $AdS_5$ spacetime. Inspired by this, many holographic models have been proposed based on either string theory constructions (top-down) \cite{Karch:2002-Adding-Flavors-To-AdSCFT-D3D7,Sakai-Sugimoto:2004} or properties of QCD (bottom-up) \cite{HardWall2005,SoftWall-01,EMD-01,Gursoy:2007-ImprovedQCD-Part1,Gursoy:2007-ImprovedQCD-Part2,Iatrakis:2010-VQCD,Danning2013-1,Danning2013-2}. The calculation of photon emission rate using gauge/gravity duality was first investigated in \cite{Caron-Huot}, where the photon emission rate of $\mathscr{N}=4$ supersymmetric Yang-Mills (SYM) plasma under both weakly and strongly coupled circumstances were given. The calculations for the non-conformal plasma soon followed up \cite{I.I.YangKiritisis:2016ugz,YangDLConductPLB,Stefano2015}. The strong electromagnetic field can be created in the HICs; consequently, extensions of photon production in anisotropic plasmas \cite{ArefevaRussia2021} have also been explored, in which an external magnetic field induced the anisotropy. Despite all these efforts, the photon emission near the CEP remains largely unexplored in the context of holography.

Among all those bottom-up models, the Einstein-Maxwell-dilaton (EMD) model \cite{EMD-01,Yang:2014bqa} was constructed to mimick the equations of state that captures the crossover transition. Recently, the machine learning method \cite{ChenXun} has been employed to determine the key parameters of the EMD model. Using this machine-learning model, we will calculate the thermal photon emission rate and electric conductivity of a strongly coupled QGP in thermal equilibrium. Actually these two quantities can be calculated using the same retarded correlator \cite{Caron-Huot}. To achieve that, we need to add a Maxwell action into the original model in the same spirit of linear-response theory, the backreaction of the newly-added action to the background should be neglected.

In this work we use the notation $d\Gamma/dk$ as photon emission rate, where $\Gamma$ denotes the number of photons emitted per unit time per unit volume and $k=\omega$ is the momentum of the photon. This follows the convention from previous holographic studies \cite{Caron-Huot,I.I.YangKiritisis:2016ugz,YangDLConductPLB}. There's also a different definition $d\Gamma/d^3k$ of photon emission rate \cite{latticePhoton-2022,HotQCDCollaboration:2024-latticePhoton,Akamatsu:2025axh}. The rest of this paper is organized as following: in Section~\ref{section2}, we give a brief review of the EMD model and introduce the machine-learning model in \cite{ChenXun}. In Section~\ref{section3-A} we first present both the deduction of DC electric conductivity and photon emission rate. We then show how the coupling function in probe action is determined in Section~\ref{section3-B}, where the electric conductivities computed at both zero and finite chemical potential are presented. Our numerical result of photon emission rate is given in Section~\ref{section3-C}. Finally, we give conclusion and discussion in Section \ref{section4}. 

\section{The Einstein-Maxwell-dilaton framework}\label{section2}
\subsection{EMD model revisit}

Firstly we give a brief introduction to the Einstein-Maxwell-dilaton (EMD) framework \cite{Gubser:2008ny,EMD-01,Danning2013-2,Yang:2015aia,Yang:2014bqa,Dudal:2017-MLsEMDsolution,Dudal:2018ztm,Fang:2015ytf,Liu:2023pbt,Li:2022erd,Critelli:2017oub,Grefa:2021qvt,ArefevaRussia2021,Arefeva:2020vae,Chen:2020ath,Chen:2019rez}, which has proven to be quite sucsessful in capturing key features of hot/dense QCD matter. Following previous studies, in the Einstein frame, the general form of the EMD action, which mimicks the QCD equation of state, takes the form 
\begin{equation}\label{EMD action}
    S=\frac{1}{16\pi G_5}\int d^5x\sqrt{-g}\left[R-\frac{f(\phi)}{4}F^2-\frac{1}{2}\partial_M\phi\partial^M\phi-V(\phi)\right].
\end{equation}
Here $G_5$ denotes the 5D Newton constant, and $R$ is the Ricci scalar. The action includes a $U(1)$ Maxwell field $A_M$ with $F_{MN}$ as its field strength tensor, a dilaton field $\phi$ and its potential $V(\phi)$. The $U(1)$ field $A_M$ is introduced as the dual part of the baryonic current, and its temporal component $A_0$ on the boundary gives the baryonic chemical potential. The five-dimensional metric $g_{MN}$ is the ($M,N=0,1,...,5$) is taken as
\begin{equation}\label{metric-ansatz}
    ds^2=\frac{L^2 e^{A(z)}}{z^2}\left[-g(z)dt^2+\frac{dz^2}{g(z)}+d\vec{x}^2\right].
\end{equation}
Here $z$ represents the holographic coordinate in the fifth dimension, and the radius $L$ of the $\rm AdS_5$ space. The dilaton field $\phi$ is introduced to break the conformal invariance.

As suggested in Refs.\cite{ChenXun,Dudal:2017-MLsEMDsolution}, the deformed metric $A(z)$ in Eq.\eqref{metric-ansatz} takes the following ansatz   
\begin{equation}
A(z)=d\ln(az^2+1)+d\ln(bz^4+1),\label{chen's ansartz for metric}\\
\end{equation}
and the parameters are fixed by employing machine-learning techniques to fit lattice QCD equation-of-state (EOS) data at zero baryon chemical potential ($\mu_B=0$). 

The gauge kinetic function $f(\phi)$ couples to the gauge field $A_\mu$ describing the chemical potential dependent property, and following \cite{ChenXun} it can be taken as 
\begin{equation}
f(\phi(z))\equiv f(z)=e^{cz^2-A(z)+k},
\end{equation}
where $c$ and $k$ are two model parameters.

The dilaton potential $V\left(\phi\right)$ can be systematically solved from the equations of motion (EoMs) 
\begin{equation}
\begin{gathered}
R_{M N}-\frac{1}{2} g_{M N} R-T_{M N}=0, \\
\nabla_M\left[f(\phi) F^{M N}\right]=0, \\
\partial_M\left[\sqrt{-g} \partial^M \phi\right]-\sqrt{-g}\left(\frac{\partial V}{\partial \phi}+\frac{F^2}{4} \frac{\partial f}{\partial \phi}\right)=0,
\end{gathered}
\end{equation}
with
\begin{equation}
\begin{aligned}
T_{M N}=\frac{1}{2}\left(\partial_M \phi \partial_N \phi-\frac{1}{2} g_{M N}(\partial \phi)^2-g_{M N} V(\phi)\right)+\frac{f(\phi)}{2}\left(F_{M P} F_N^P-\frac{1}{4} g_{M N} F^2\right) .
\end{aligned}
\end{equation}

In the following calculation, we set the AdS radius $L=1$. In the machine-learning EMD model \cite{ChenXun}, the specific expressions of the fuctions mentioned above are given by:
\begin{align}    
    g(z)&=1-\frac{1}{\int_0^{z_h}dxx^3e^{-3A(x)}}\left[\int_0^zdxx^3e^{-3A(x)}+\frac{2c\mu^2e^k}{(1-e^{-cz_h^2})^2}\det \mathcal{G}\right], \label{BH factor}\\
    \det\mathcal{G}&=
    \begin{vmatrix}
    \int_0^{z_h}dyy^3e^{-3A(y)}&\int_0^{z_h}dyy^3e^{-3A(y)-cy^2}\\
        \int_{z_h}^{z}dyy^3e^{-3A(y)}&\int_{z_h}^zdyy^3e^{-3A(y)-cy^2}
    \end{vmatrix},\\
    \phi'(z)&=\sqrt{6(A'^2-A''-2A'/z)},\\
    A_t(z)&=\mu\frac{e^{-cz^2}-e^{-cz^2_h}}{1-e^{-cz^2_h}},\\
    V(z)&=-3z^2ge^{-2A}\left[A''+A'(3A'-\frac{6}{z}+\frac{3g'}{2g})-\frac{1}{z}(-\frac{4}{z}+\frac{3g'}{2g})+\frac{g''}{6g}\right].
\end{align}

The Hawking temperature is given by:
\begin{equation}\label{temp}
    T=\frac{z_h^3e^{-3A(z_h)}}{4\pi\int_0^{z_h}dyy^3e^{-3A(y)}}[1+\frac{2c\mu^2e^k(e^{-cz_h^2}\int_0^{z_h}dyy^3e^{-3A(y)}-\int_0^{z_h}dyy^3e^{-3A(y)-cy^2})}{(1-e^{-cz_h^2})^2}]
\end{equation}
and the baryon chemical potential $\mu=3\mu_q$ correponds to 
the ultraviolet (UV) boundary ($z=0$) condition
\begin{eqnarray}
A(0)&=&-\sqrt{\frac{1}{6}} \phi(0) =0,  g(0)=1,  \\
A_t(0)&=&\mu+\rho^{\prime} z^2+\cdots.
\end{eqnarray}
Here $\rho^{\prime}$ corresponds to a quantity proportional to the baryon number density.  Following Refs.\cite{Critelli:2017oub,Zhang:2022uin}, the baryon number density and and baryon chemical potential can be derived as
\begin{eqnarray}
    \rho  &=&\left|\lim _{z \rightarrow 0} \frac{\partial \mathcal{L}}{\partial\left(\partial_z A_t\right)}\right| 
 = -\frac{1}{16\pi G_5} \lim _{z \rightarrow 0}\left[\frac{\mathrm{e}^{A(z)}}{z} f(\phi) \frac{\mathrm{d}}{\mathrm{d} z} A_t(z)\right],\\
\mu&=&-\sqrt{\frac{-1}{\int_0^{z_h} y^3 e^{-3 A} d y \int_{y_g}^y \frac{x}{e^A f} d x}} \int_0^{z_h} \frac{y}{e^A f} d y.
\end{eqnarray}

All parameters have already been determined by a machine-learning algorithm \cite{ChenXun} and are listed in Table~\ref{parameters} for the $N_f=2+1$ case. With these parameters we will investigate the electromagnetic properties of the holographic QGP phase. 
\begin{table}[htbp]
    \centering
        \begin{tabular}{|c|c|c|c|c|c|c|}
        \hline
        $N_f$ & a & b & c & d & k & $G_5$\\
        \hline
        2+1 & 0.204 & 0.013 & -0.264 & -0.173 & -0.824 & 0.400\\
        \hline
    \end{tabular}
    \caption{Parameters for $N_f=2+1$ system from machine-learning \cite{ChenXun}.}
    \label{parameters}
\end{table}

\subsection{Probe action}

As can be seen, the EMD action given by Eq.\eqref{EMD action} does not contain any term involving the electromagnetic current. However, such a term is necessary to study the electromagnetic properties of the dual medium. Based on the previous calculations of thermal photon emission rate and transportation coefficients \cite{Stefano2013,Stefano2015,Caron-Huot}, we take the same method to add a probe action to \eqref{EMD action}. In the same spirit of linear response theory, the newly added action only serves as an external probe and its back-reaction to the background metric is neglected. This in fact corresponds to the case where the electric charge chemical potential vanishes, i.e., $\mu_Q=0$. Building on previous studies, the probe action can be simply taken as 
\begin{equation}\label{probe action}
    S_{EM}=-\frac{1}{16\pi G_5}\int d^5x\sqrt{-g}\frac{f_Q(\phi)}{4}F_Q^2.
\end{equation}
Here the additional $U(1)$ gauge field $A_Q$, whose field strength is $F_Q$, is holographically dual to the electromagnetic current, and the additional coupling function $f_Q(\phi)$ characterizes the coulping of this additional gauge filed to the dilaton. We will compare the following two choices for $f_Q(\phi)$ in later calculations

\begin{align}
    &(1):~f_Q(\phi)=\alpha_1=\mathrm{constant}\label{constant fQ},\\
    &(2):~f_Q(\phi)=\alpha_1+\alpha_2 \mathrm{sech[\beta\phi(z)]}\label{my good fQ},
\end{align}
with parameters $\alpha_1,\alpha_2, \beta$. The constant choice can be treated as a special case of \eqref{my good fQ}, where $\alpha_2=0$, $f_Q(\phi)$ is fully parameterized by $\alpha_1$. 

These two forms of $f_Q$ are motivated by the following considerations. In Refs.~\cite{Stefano2013,Stefano2015},  $f_Q(\phi)$ is taken to be either a single $\mathrm{sech(\phi)}$ or a linear combination of two such terms. Here the constant $\alpha_1$ in \eqref{my good fQ} is introduced  so that the DC conductivity saturates quickly enough at high temperature. Besides, one might wonder why we use a constant $f_Q(\phi)$ as in \eqref{constant fQ}. Actually, using gauge/gravity duality, when one of the earliest calculations of thermal photon emission rate was carried out \cite{Caron-Huot}, the coupling of the probe action is a constant. Also, in several later studies, the $f_Q(\phi)$s  used are all monotonically descreasing in $z\in [0,z_h]$ from small positive values to zero with a slow speed \cite{Stefano2013,Stefano2015,YangDLConductPLB}. Specifically the plots of these choices are relatively flat and look like the ones of a constant, thus we regard the \eqref{constant fQ} choice as not far from a reasonable one. And actually, it'll be shown that the constant choice \eqref{constant fQ} does give a result that resembles the $\phi(z)$-dependent choice \eqref{my good fQ}.

As for the model parameters, we expect our thermal photon emission rate to reach the result of $\mathscr{N}=4$ SYM plasma \cite{Caron-Huot} at high temperature (above $3T_c$), and since the overall amplitude of thermal photon emission rate is proportional to the electric conductivity, our choice of $f_Q(\phi)$ should consequently coincide with the electric conductivity of $\mathscr{N}=4$ SYM plasma as well at high $T$. Therefore, we shall adjust the parameters in our coupling based on the requirement that our electric conductivity matches the SYM result and lattice calculations.

Finally, we emphasize that since the baryonic number has already been included in the EMD action \eqref{EMD action}, it is straightforward to consider the electromagnetic properties of the dual medium at finite densities. This is particularly relevant at the critical end point (CEP), which has not yet been carefully analyzed using holographic models.

\section{Thermal photon emission rate, electric conductivity calculation}\label{section3}

\subsection{Derivation of thermal photon emission rate and electric conductivity}\label{section3-A}
In this section, we describe the methods for calculating the thermal photon emission rate and the DC electric conductivity using holography. To study thermal photon emission, we adopt the static background metric described in Sec.~\ref{section2}, assuming that the plasma is in local thermal equilibrium in this stationary space-time. In this case, an early work \cite{Caron-Huot} gave the thermal photon emission rate and severval transport coefficients of $\mathcal{N}=4$ Super Yang-Mills plasma based on gauge/gravity duality, and here we take similar procedures. The number of photons emitted per unit time per unit volume is denoted by $\Gamma$ and it takes the following form
\begin{align}
    d\Gamma&=\frac{d^3k}{(2\pi)^3}\frac{e^2}{2|\mathbf{k}|}\eta^{\mu\nu}C_{\mu\nu}^{<}(K)|_{k^0=\omega=|\mathbf{k}|}\label{dGammad3k},\\
    C_{\mu\nu}^{<}(K)&=n_b(\omega)\chi_{\mu\nu}(K)=\frac{1}{e^{\omega\beta}-1}\chi_{\mu\nu}(K),\label{equilibrium wightman and chi}
\end{align}
where $C_{\mu\nu}^{<}(K)$ is the Wightman function of th electromagnetic currents
\begin{equation}
    C_{\mu\nu}^{<}(K)=\int d^4 Xe^{-K\cdot X}\langle J^{EM}_\mu(0)J^{EM}_\nu(X)\rangle,
\end{equation}
and $n_b(\omega)$ denotes the Bose-Einstein distribution function and $\beta=1/T$, while $\chi_{\mu\nu}(K)$ is the the spectral density and \eqref{equilibrium wightman and chi} holds in the case of thermal equilibrium. Here, $X=(t,\mathbf{x})$ and $K=(\omega,\mathbf{k})$ denote spacetime coordinates and momentum, the capital letters represent 4-vectors, while the bold form like $\mathbf{k}$ and the ones with latin index $i$ like $k^i$ denote 3-vectors.

For $\chi_{\mu\nu}(K)$, it's proportional to the imaginary part of the retarded current-current correlator $C^{ret}_{\mu\nu}(K)$:
\begin{equation}
   \chi_{\mu\nu}(K)=-2\mathrm{Im}C_{\mu\nu}^{ret}(K).
\end{equation}
When $T\ne 0$, we can separate $C^{ret}_{\mu\nu}(K)$ into its transverse ($\Pi^T$) and longitudinal ($\Pi^L$) part :
\begin{equation}
    C^{ret}_{\mu\nu}(K)=P^T_{\mu\nu}(K)\Pi^T(\omega,\mathbf{k})+P_{\mu\nu}^L(K)\Pi^L(\omega,\mathbf{k}),
\end{equation}
where $P^T_{\mu\nu}(K)$ and $P_{\mu\nu}^L(K)$ are transverse and longitudinal projectors:
\begin{align}
    P_{00}^T(K)=0,\quad P_{0i}^T(K)=0,\\
    P_{ij}^T(K)=\delta_{ij} - k_i k_j/\mathbf{k}^2,\\
    P_{\mu\nu}^L(K) \equiv P_{\mu\nu}(K) - P_{\mu\nu}^T(K),\\
    P_{\mu\nu}(K)=\eta_{\mu\nu} - K_\mu K_\nu/K^2.
\end{align}
Here $\eta_{\mu\nu}$ denotes the Minkowski metric $(-,+,+,+)$. Contracting the indices of $\chi_{\mu\nu}$, one can get:
\begin{equation}
     {\chi^\mu}_{\mu}(K)=-4\mathrm{Im}\Pi^T-2\mathrm{Im}\Pi^L.
\end{equation}
For photons, the longitudinal part $\mathrm{Im}\Pi^L$ vanishes. Therefore we only need the transverse part for the following calculations.

Now, with some simplication to \eqref{dGammad3k} (see appendix \ref{Appendix} for more details), we can finally express the photon emission rate $d\Gamma/dk$ in terms of $\mathrm{Im}\Pi^T$:

\begin{align}
    \frac{d\Gamma}{dk}=-4\frac{\alpha_{EM}}{\pi}k\frac{1}{e^{k\beta}-1}\mathrm{Im}\Pi^T,\label{emission rate}
\end{align}
where the fine structure constant $\alpha_{EM}\equiv\frac{e^2}{4\pi}$ whose value is taken as $\frac{1}{137}$ in the following calculations. Now it becomes clear that to get the emission rate $\frac{d\Gamma}{dk}$, one only need to calculate the correlation function $C_{\mu\nu}^{ret}(K)$ (or it's transverse part $\Pi^T$) and substitute it back into \eqref{emission rate}.

On the other hand, the calculation of electric conductivity $\sigma_Q$ can be expressed in terms of the same current-current correlation function \cite{Caron-Huot}:
\begin{align}
    \sigma_Q=\lim_{\omega\to 0}\frac{e^2}{4T}\eta^{\mu\nu}C^{<}_{\mu\nu}(K)|_{k^0=\omega=|\vec{k}|},\label{C-H's conduct}
\end{align}
and it can be further simplified as:
\begin{align}
    \sigma_Q=-\lim_{\omega\to 0} 4\pi\alpha_{EM}\mathrm{Im}\Pi^T\frac{1}{\omega}.\label{final expression for conduct}
\end{align}
Thus, once we obtain the transverse part of retarded correlator $\Pi^T$, we can get both the $\sigma_Q$ and photon emission rate $d\Gamma/dk$. According to the AdS/CFT dictionary, the generating functional of the 4-d gauge theory can be expressed in terms of the on-shell action of the 5-d gravitional theory. Following the prescription in \cite{DTS-ADS-Prescript}, one can eventually get the transverse part of retarded correlation function (for details, please see appendix \eqref{Appendix}):
\begin{equation}\label{on shell action final}
    \Pi^T=\lim_{z\to 0}\left[-2\cdot\frac{1}{16\pi G_5}g^{zz}\sqrt{-g}\frac{f_Q(\phi)}{4}\frac{\partial_z A_i(z,k)}{A_i(z,k)}\right],
\end{equation}
where we choose the 4-momentum of photon to be $K=(\omega,0,0,k)$, and the index $i$ should take the value of 1 or 2 (the summation over $i$ is not implied here). The overall factor "2" here comes from the prescription. With \eqref{on shell action final}, once the equation of motion is solved, one can easily obtain the electric conducivity and photon emission rate. In this paper, our choice of coupling doesn't provide us with an analytical solution, so we solve the equation of motion using a numerical method. Moreover, to obain a physical solution, the incoming boundary condition should be imposed at the horizon.
\subsection{Numerical result of electric conductivity}\label{section3-B}

\subsubsection{Elctric conductivity at zero chemical potential}

\begin{figure}[H]
\centering
\includegraphics[width=0.4\textwidth]{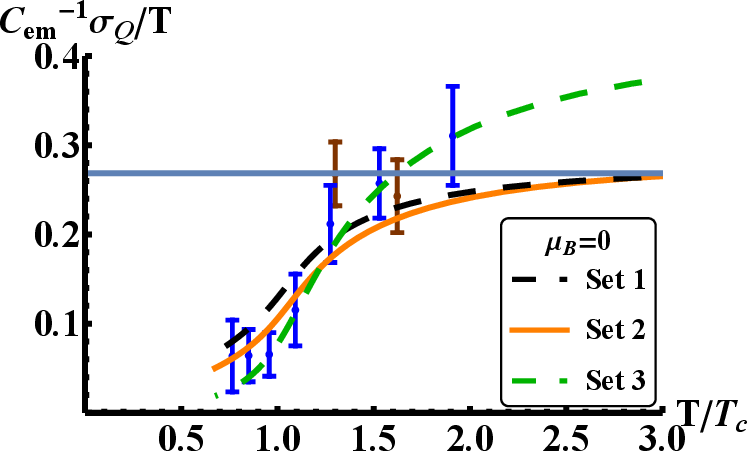}
\caption{\label{conduct and fQ at 0mu} Holographic results of DC electric conductivity scaled by $T C_{em}$ as a function of $T/T_c$ for three different coupling functions $f_Q(\phi)$, comparing with lattice results. The lattice results are marked by brown \cite{2019ELEconductivityLattice} and blue \cite{GertAartsAmato2013PRL,GertAartsAmato2014JHEP} with error bars, while the horizontal lightblue line corresponds to the $\mathcal{N}=4$ SYM result \cite{Caron-Huot}. The other three curves represent holographic results of electric conductivity using three different sets of parameters: 1) black dashed line for $f_Q=\alpha_1=1.19$; Orange solid line for $\alpha_1=0.81,\alpha_2=0.41,\beta=0.4$; Green dashed line for $\alpha_1=0,\alpha_2=1.8,\beta=0.4$. We also normalize the temperature using (pseudo)transition temperature $T_c$ at zero chemical potential.}
\end{figure}

Given the metric described in Sec.~\ref{section2} and specific values of the parameters in $f_Q$, it is straightforward to obtain the holographic results for the DC electric conductivity using the formula derived in the previous section. We calculate the electric conductivity at $\mu_B=0$ using three sets of parameters listed in Table~\ref{parameters for fQ}. 
The numerical results from holographic model for DC electric conductivity scaled by $T C_{em}$ are shown in Fig.\ref{conduct and fQ at 0mu} as a function of $T/T_c$, comparing with lattice results \cite{2019ELEconductivityLattice} and  \cite{GertAartsAmato2013PRL,GertAartsAmato2014JHEP} shown in data with error bars, where the blue horizontal line is the $\mathcal{N}=4$ SYM limit (at $N_c=3$) from \cite{Caron-Huot},  three curves correspond to holographic results with three different set of parameters, respectively. Here $C_{em}=e^2\sum_fq_f^2$, $q_f^2$ refers to the number of electric charge carried by the quark, and in the case of $N_f=3$ one has $C_{em}=2e^2/3$. 
\begin{table}[htbp]
    \centering
        \begin{tabular}{|c|c|c|c|}
        \hline
        Parameters & $\alpha_1$ & $\alpha_2$ & $\beta$\\
        \hline
        Set 1      & 1.19       & 0          & 0      \\
        \hline
        Set 2      & 0.81       & 0.41       & 0.4    \\
        \hline
        Set 3      & 0          & 1.8        & 0.4    \\
        \hline
    \end{tabular}
    \caption{Three different sets of parameters for $f_Q(\phi(z))$.  Set 1 and Set 2 are chosen to make electric conductivity reach SYM limit at high temperature above $3T_c$, while Set 3 provides a better match to the lattice data within the error bars.}
    \label{parameters for fQ}
\end{table}

The lattice results \cite{GertAartsAmato2013PRL,GertAartsAmato2014JHEP,2019ELEconductivityLattice} have different values of $T_c$ at zero chemical potential, $T_c$ is around 184 MeV in \cite{GertAartsAmato2013PRL,GertAartsAmato2014JHEP} and 153 MeV in \cite{2019ELEconductivityLattice}. Thus for each type of data, we use the normalized temperature $T/T_c$. For the machine-learned model, $T_c=$128 MeV when $\mu_B=0$. In Fig.\ref{conduct and fQ at 0mu}, the lattice data exhibits an increase in electric conductivity with temperature, and it matches the $\mathcal{N}=4$ SYM result at high temperature. It is observed that for Set 1 and Set 2, the DC electric conductivity scaled by $T C_{em}$ grows fast in the region of $1\sim 1.5T_c$, and slower above $1.5T_c$ before reaching the SYM limit.

Behaviors of electric conductivity can be modified by different parameters. Set 1 and Set 2 enable the electric conductivity to reach SYM limit at high temperature (above $3T_c$). If we relax the high-temperature constraint, a better fit to the lattice data can be achieved using Set 3. The result is shown by the green dashed curve in Fig.\ref{conduct and fQ at 0mu}.
Using parameters in Set 1 and Set 2 the resulting electric conductivity reproduces the key qualitative trends seen in lattice QCD and yields a semi-quantitative agreement with existing data. As discussed before, the constant coupling (black dashed line) does give a result close to the lattice simulation. But it still deviates from the lattice error bars slightly at low $T$, also it lacks flexibility for further adjustment. Meanwhile, although Set 2 (orange solid line) is very close to the constant one, it is more adjustable.

However, due to the broad error bars of the lattice result, it's always possible to find many couplings by either changing parameters or using other forms such that whose outcomes appear in "good" agreement with the lattice data. Rather than fine-tuning the fit within the large lattice error bars, we focus on the generic behavior of the photon emission rate and the electric conductivity across the transition temperature. We therefore choose Set 2 in the following calculations, whose photon emssion rate also approaches the $\mathcal{N}=4$ SYM value at high temperatures and remains saturated thereafter. An analogous prescription was adopted in Ref.~\cite{YangDLConductPLB}, both this choice and SYM calculations were found to be close to experimental data \cite{I.I.YangKiritisis:2016ugz}. We thus consider it as a reasonable choice.

\subsubsection{Electric conductivities at dfferent chemical potentials}

\begin{figure}[H]
    \centering
    \begin{tabular}{cc}
        \includegraphics[width=0.4\textwidth]{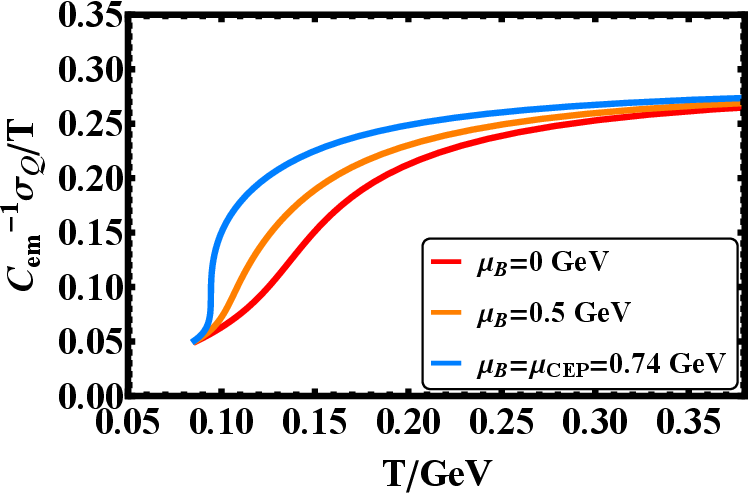}
		\includegraphics[width=0.4\textwidth]{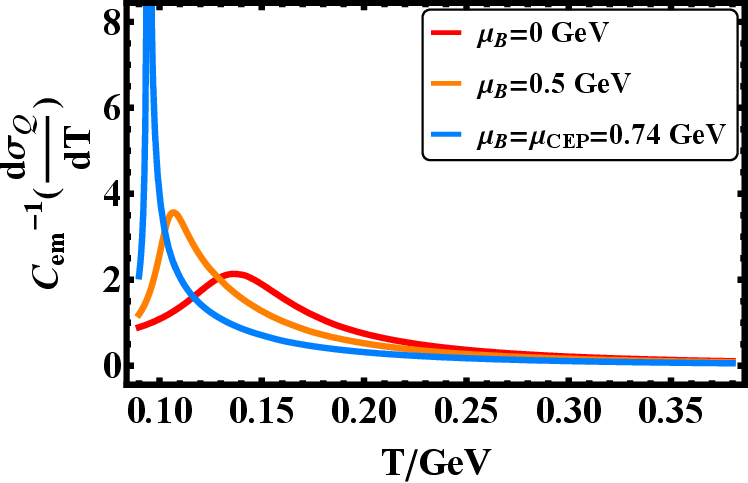}
	\end{tabular}
    \caption{\label{conductivity and 1st order derivatives} \textbf{Left}: Electric conductivitis calculated using parameters in Set 2 as a function of the temperature for $\mu_B=0$ (red), $0.5$ (orange) and $0.74$ (blue) GeV. \textbf{Right}: First order derivatives of electric conductivity at $\mu_B=0$ (red), $\mu_B=0.5$ GeV (orange) and $\mu_B=\mu_{\mathrm{CEP}}=0.74$ GeV. Each curve exhibits a peak which corresponds to the inflection point in the left pannel.}
\end{figure}

Since the model parameters are already fixed, finite-density effects follow directly from the holographic setup without further tuning. We take $\mu_B=0,0.5,0.74$ GeV as examples, the electric conductivities (left) and their first order derivatives (right) are shown in Fig.~\ref{conductivity and 1st order derivatives}. The red, orange and blue curves correspond to $\mu_B=0,0.5,0.74$ GeV cases respectively. The machine-learning model predicts the CEP to be at $T_{CEP}=0.094$ GeV, $\mu_{CEP}=0.74$ GeV, so the blue line is expected to capture the CEP-induced critical behavior. In Fig.~\ref{conductivity and 1st order derivatives} (left), all curves monotonically increases with respect to $T$ and $\mu_B$, they are convex below $T_c$ and becomes concave above $1.5T_c$, indicating the presence of an inflection point around $T_c$. This can been seen more clearly from the right pannel, where first order derivatives are non-negative and approach zero at large $T$. The exact positions of inflection points are determined by the locations of the peaks in the derivative plot. The locations of the inflection points are given in Table.\ref{inflection points and Tcs}.

The most rapid growth of electric conductivity occurs at the inflection points thus a correspondingly steep rise in the photon emission rate is expected, as both quantities are governed by the same current-current correlator. When increasing chemical potential, the conductivity curve becomes steeper, eventually at CEP, the conductivity curve develops into a singular slope. From the perspective of the first-order derivative, the width of the peak shrinks and its height rises with increasing chemical potential, while the singular slope at the CEP corresponds to a divergence of the first-order derivative. At high temperature, the increase of all three curves slow down and stop at certain values. Specifically, at $T=0.35$ GeV, the normalized conductivity ${C_\mathrm{EM}}^{-1}\sigma_Q/T$ of $\mu_B=0.74$ GeV reaches 0.272 ($\mu_B=0.74$), while for $\mu_B=0$ and $\mu_B=0.5$ GeV case, it's 0.261 and 0.266 respectively.

\begin{table}[H]
    \centering
        \begin{tabular}{|c|c|c|c|}
        \hline
        $\mu_B$(GeV) & 0 & 0.5 & 0.74\\
        \hline
        Transition Temperature(GeV) & 0.128 & 0.103 & 0.094 \\
        \hline
        Inflection Points(GeV) & 0.136 & 0.107 & 0.094\\
        \hline
    \end{tabular}
    \caption{The position of inflection points and transition temperatures are given. In the crossover region, the $T_c$s are very close to the inflection points, and at the CEP, the inflection point coincides with the transition temperature.}
    \label{inflection points and Tcs}
\end{table}

For $\mu_B=0,0.5$ GeV, our results for the inflection points are very close to the $T_c$ given by the machine-learning EMD model (difference within 10 MeV). As mentioned before, in the crossover region ($\mu_B<\mu_{CEP}$), the $T_c$ is in fact a pseudo-critical temperature since the transition there is a continous type. In lattice calculations, different observables often lead to different values of $T_c$, and since the machine learning EMD model we used is based on lattice result, we do not expect the inflection points of our electric conductivity to match the $T_c$ perfectly. However, such a small difference still indicates the relatively rapid rise of electric conductivity occurs when the phase transition happens. On the other hand, the coincidence of the inflection point with transition temperature is exact at the CEP, which means that the most significant growth of DC electric conductivity will occur at the CEP. 

Apart from these features, one might have noticed that the conductivity curves seem to coincide at low temperatures, and this has to do with the metric at different chemical potentials. As shown from the deduction of $\sigma_Q$ in section \ref{section4}, the key step to obtain DC conductivity is to substitute the solution of equation of motion into the action. In this step the only difference between all three cases lies in the metric. According to \eqref{metric-ansatz},\eqref{chen's ansartz for metric} and \eqref{BH factor}, only the blackening factor \eqref{BH factor} in the metric depends on $\mu_B$ and $z_h$. The influence of $\mu_B$ on metric can thus be investigated through the temperature-horizon plot. 

\begin{figure}[H]
\centering
\includegraphics[width=0.4\textwidth]{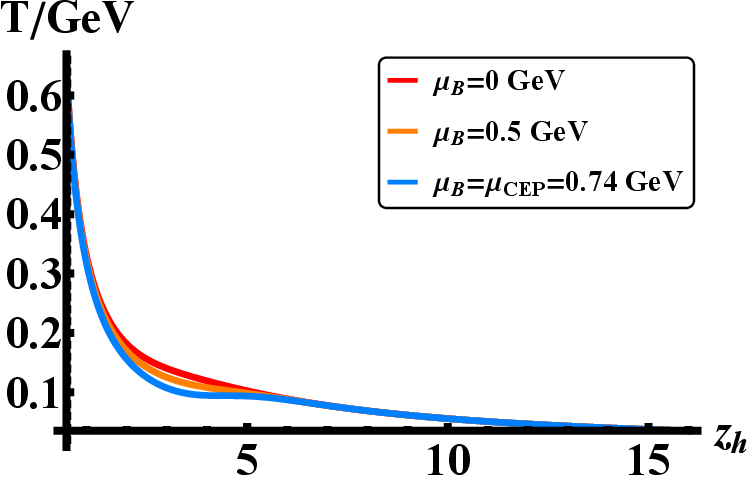}
\caption{\label{T-zh} Temperature dependence on $z_h$ at $\mu_B=0$ (red), $0.5$ GeV (orange) and $0.74$ GeV (blue). At both high and low temperature, all three curves converge into a single one. We only consider the region of midium temperature where the temperature changes more significantly with $\mu_B$.}
\end{figure}

In Fig.\ref{T-zh} we plot the temperature against the position of horizon $z_h$ for different values of the chemical potential $\mu_B$. The red, blue, orange curves represents the temperature dependence on $z_h$ at $\mu_B=0$, $\mu_B=0.5 \rm{GeV}$, $\mu_B=\mu_{\rm{CEP}}=0.74\rm{GeV}$ respectively. For both very small and large $z_h$, corresponding to very large and small temperatures, the three curves converge into a single one. Under this circumstance, all three curves correspond to the same value of $z_h$ for a certain temperature. This implies that the differences in the metric are negligible in these regions. Consequently, the values of $\sigma_Q$ coincide at very low and high temperatures, as shown in Fig.\ref{conductivity and 1st order derivatives}. However, we would like to point out that  this coincidence at low temperature might be due to some model artifact . In fact, this model is reliable only within a certain temperature range, where $T$ is neither too high nor too low. Finally, Fig.\ref{T-zh} shows that the CEP in this EMD model can be attributed to the flat shape of the  $T-z_h$ curve near the CEP. Consequently, the divergence of the temperature derivative may also manifest in many other thermal quantities. Thus, we expect that this divergence is independent of the holographic settings chosen here.

\subsection{Numerical result of thermal photon emission rate}\label{section3-C}
\begin{figure}[H]
    \centering
    \begin{tabular}{cc}
		\includegraphics[width=0.98\linewidth]{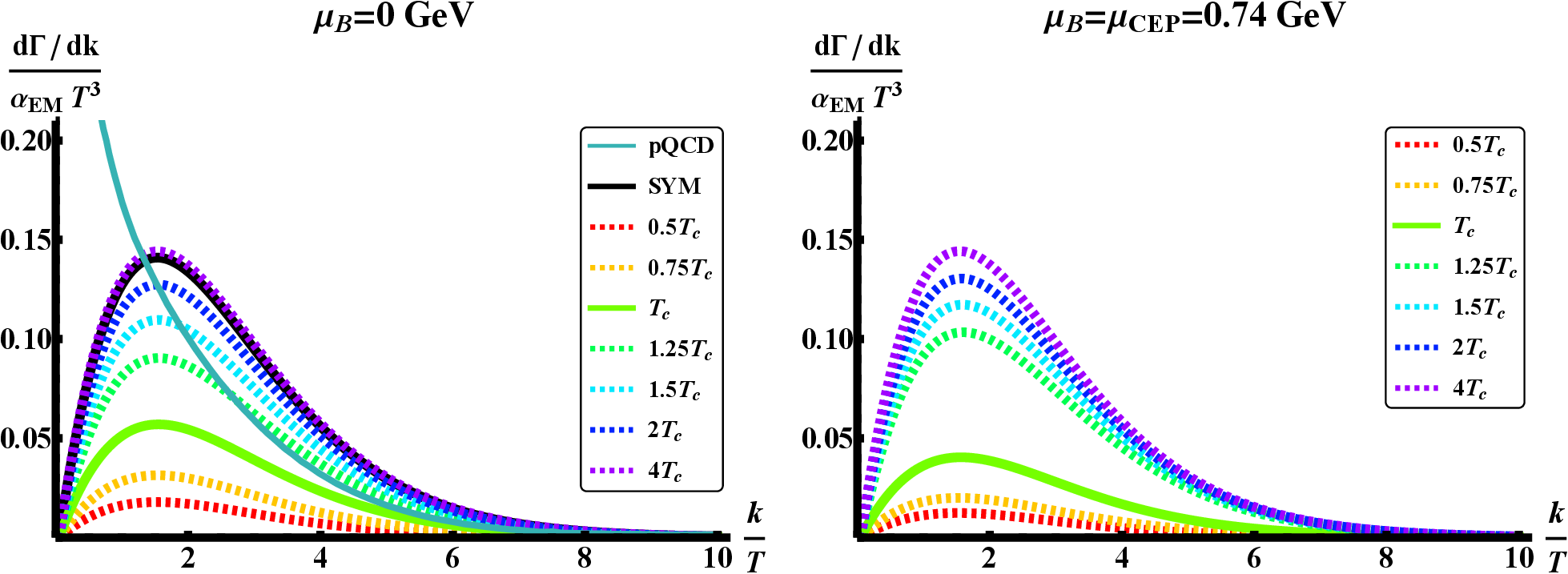}
	\end{tabular}
    \caption{\label{2D EmissionRate at 0 and CEP} 
    Thermal photon emission rate plotted agaginst the momentum $k/T$, divied by fine structure constant $\alpha_\mathrm{EM}$ and temperature cubed. \textbf{Left}: $\mu_B=0$. The dashed lines represent thermal photon emission rate calculated using parameter Set 2 at different temperatures. From red to purple, the temperature increases. The SYM and pQCD (upto NLO) result are also shown for comparison, they are plotted using black and light blue solid line respectively. \textbf{Right}: $\mu_B=\mu_\mathrm{CEP}=0.74$ GeV case. The coupling function is also parameterized by Set 2. The emission result computed at CEP is plotted by the green solid curve.}
\end{figure}

\begin{figure}[H]
    \centering
    \includegraphics[width=0.74\linewidth]{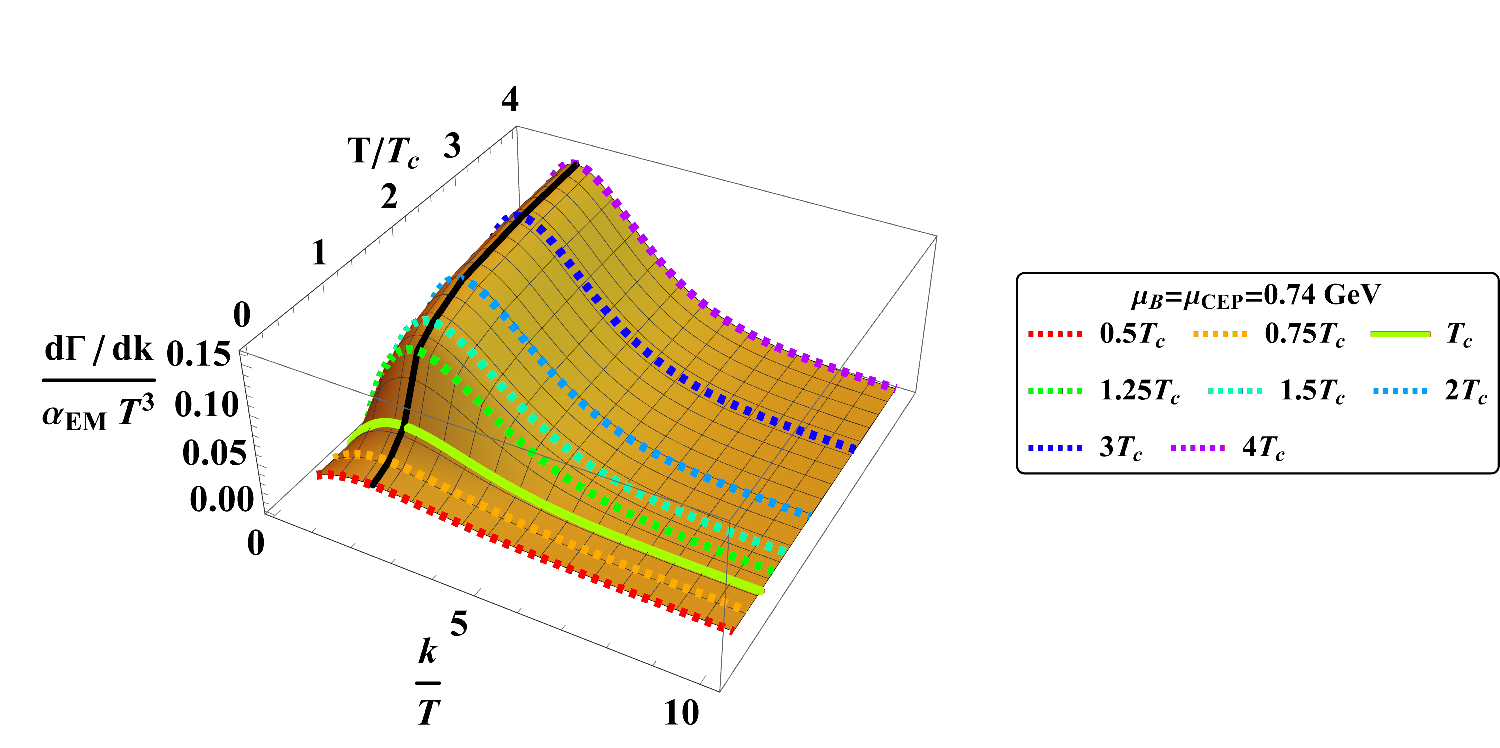}
    \caption{Photon emission rate at $\mu_B=\mu_{\mathrm{CEP}}=0.74$GeV. Each curve corresponds to one particular temperature. From red to purple, the corresponding temperatures rise from $0.5T_c$ to $4T_c$. In particular, the green solid line represents the photon emission rate at CEP. The overall amplitude of  photon emission rate increases with temperature. Similar to the conductivity before, the most rapid growth occurs at the critical temperature as shown by the obvious slope in this 3D figure. In order to show the slope tendency more clearly, a black line connecting the peaks of all curves is depicted.}
    \label{3D Emission rate at CEP}
\end{figure}

Similar to the conductivity, the photon emission rate can also be obtained from the current-current correlator. In Fig.\ref{2D EmissionRate at 0 and CEP},  we show the normalized photon emission rate $\frac{d\Gamma/dk}{\alpha_{EM}T^3}$ against the momentum of photon $k/T$ (nomalized by temperature). The analytic expression for the SYM and pQCD result are given in \cite{Caron-Huot} and \cite{pQCD-NLO2013} respectively, for the SYM emission rate we take $N_c=3$. With our normalization notation, both SYM and pQCD emssion rate $\frac{d\Gamma/dk}{\alpha_{EM}T^3}$ are functions of $k/T$ only and independent of temperature. In Fig.\ref{2D EmissionRate at 0 and CEP}, the pQCD result \cite{pQCD-NLO2013} corresponds to the $g_s=2$ and $N_f=N_c=3$ case, whose 
temperature is estimated to be over $2$ GeV ($\Lambda_{QCD}=200$MeV). At large $k/T$ region, our result resembles the pQCD calculation, while at small $k/T$ region, pQCD result deviates from ours greatly and eventually diverges at $k/T=0$.

In this figure, the left pannel shows the result of zero chemical potential while the right one respresents the $\mu_B=\mu_{\mathrm{CEP}}$ case. In both plots, each curve corresponds to a certain temperature, they are chosen to be $T=0.5,\ 0.75,\ 1,\ 1.25,\ 1.5,\ 2,\ 4$ $T_c$ (the first 5 temperatures increases linearly). The emission rate at CEP is marked by a solid green line. Our results share some features with those from previous holographic studies, which include:
\begin{enumerate}
    \item 
    For each curve the normalized photon emission rate first increases and then falls with respect to $k/T$ (momentum of photon normalized by temperature). The maximum value of the normalized emission rate is located around $k=1.5T$ \cite{I.I.YangKiritisis:2016ugz} (for our result, maximum values sit at $k=1.55T$).
    \item 
    The overall amplitude of $\frac{d\Gamma/dk}{\alpha_{EM}T^3}$ monotonically increases as one increase the temperature or $\mu_B$ (as in \cite{Stefano2015,I.I.YangKiritisis:2016ugz}), at high-enough $T$ it eventually saturates and almost stops at a certain distribution (as in \cite{I.I.YangKiritisis:2016ugz,YangDLConductPLB}). With our choice of $f_Q(\phi)$, $\frac{d\Gamma/dk}{\alpha_{EM}T^3}$ reaches the $\mathscr{N}=4$ SYM result \cite{Caron-Huot} at high $T$, and eventually saturates at a value slightly higher than SYM result.
\end{enumerate}

Apart from these common features, in the left pannel of Fig.~\ref{2D EmissionRate at 0 and CEP}, one can observe that the emission rate at $T=T_c$ is relatively far from its neighboring curves ($T=0.75T_c$, $1.25T_c$). Further more, the low temperature curves ($T=0.5T_c$ and $T=0.75T_c$) and high temperature curves ($T=2T_c$ and $T=4T_c$) are rather closer to each other. This implies that the emission rate increases faster around $T_c$, while at both relatively lower ($<0.8T_c$) or higher ($>2T_c$) temperature, the increase of $\frac{d\Gamma/dk}{\alpha_{EM}T^3}$ is much slower. In the right pannel($\mu_B=\mu_{\mathrm{CEP}}$), the same feature becomes more obvious.

This behavior can be checked more easily in Fig.\ref{3D Emission rate at CEP}, where both temperature and momentum dependence of emission rate are shown. Same to Fig.\ref{2D EmissionRate at 0 and CEP}, each colored curve respresents emission rate at a specific temperature. In Fig.\ref{3D Emission rate at CEP} the $x$ and $y$ axes denote momentum and temperature respectively, and the $z$-direction stands for the emission rate. Result at CEP is marked by a solid green line. The fast growth of emission rate is characterized by the slope in Fig.\ref{3D Emission rate at CEP}. To see how the emission rate varies according to the temperature more clearly, a black line connecting the maximum values of the emission rate is also shown in Fig.\ref{3D Emission rate at CEP}. Actually, from the deduction one can see that the photon emission rate  depends on $z_h$ continuously. Therefore, the divergence of $d z_h/dT=1/(d T/dz_h)$ at CEP might cause the divergence of the photon emission rate, since $d/dT=d/d z_h \times dz_h/dT$. Therefore, we also expect a divergence in the temperature derivative of the photon emission rate at the CEP, which can be considered as a generic prediction from the EMD model. Since the qualitative results of the temperature derivative are the same as those of electric conductivity, which originate from the same correlators, we do not show the plot of the temperature derivative here. 

The significant difference between our result and the ones from pQCD is not surprising, since pQCD method is reliable only at sufficiently high temperature where the QGP is weakly coupled. On the contrary, EMD model was constructed to capture the feature of a strongly coupled QGP and is thus limited to a lower temperature range. Therefore one should not expect that the result obtained by this model can be continuously extended to weakly coupled regimes. Moreover, as mentioned above, our result saturates at a certain distribution at high enough temperatures. However a true holographic dual theory to QCD should be able to reproduce the pQCD predictions. This suggests that for the emission rate given by such a complete dual theory, the peak in the plot should increase in height and shift towards the low $k/T$ region as the temperature rises. We will leave this task to the future.

\section{Concluding Remarks}\label{section4}
In a machine-learning holographic EMD model with a CEP at finite baryon density, we calculate the thermal photon emission rate and the DC electric conductvity in thermal equilibrium at different temperatures and baryonic chemical potentials, especially around the CEP. We choose the overall normalization factor to make our thermal photon emission rate reach the SYM limit above $3T_c$. Our result on photon emission rate captures some common features of previous holographic calculations, and our calculation shows that the thermal photon emission rate experiences the most rapid growth at the CEP with the increase of temperature. Moreover, in previous works, the eletric conductivity is described to have a "jump" or a "rapid growth" around the transition temperature. In this work, we explore the behavior of eletric conductivity around the CEP carefully. It is found that, within the model we used, the eletric conductivity reaches its inflection point around the transition temperature in the crossover region (with a difference less than 10 MeV). It is worthy of mentioning that the temperature derivative of both the DC eletric conductivity and photon emission rate show divergence at the CEP which corresponds to a singular slope at $\mu_B=\mu_{CEP}$. One might ask whether the choice of $f_Q(\phi)$ affects our result, in fact as we pointed out in \ref{section2}, the $f_Q(\phi)$s usually vary slowly in the region we care, therefore we consider our conclusion as a general one and changing $f_Q(\phi)$ shouldn't make the transition temperature deviate too much from the inflection point of $\sigma_Q/T-T$ plot. We regard the singular slope of DC electric conductivity as a signature of the CEP. 

The behaviors of photon emission rate and electric conductivity are not unanimous for different models. Although our result matches other previous holographic and lattice result, another recent paper using NJL model \cite{Nishimura:2023-diverge-conductivity} gave different predictions of DC electric condcutivity. In their work, instead of a significant increase, a divergence in DC electric conductivity is expected to occur at the CEP. A possible mechanism for this enhancement might be due to the non-equilibrium properties near the critical point \cite{Akamatsu:2025axh}. These properties cannot be captured in our near-equilibrium analysis. Therefore, although the current analysis has already provided partial enhancement of the photon emission rate near the CEP (the divergence of the growth rate), further investigations, especially including coupling with the critical long-wavelength modes, are still needed to achieve a full understanding through holography. However, since the electric conductivity and photon emission rate are response of the vector-current from the system, not directly couple to the scalar order parameter, therefore it might be reasonable that
these quantities do not show divergences at CEP, but their temperature derivative divergence captures some feature of CEP. It deserves further studies in the future to clarify the feature of the electric conductivity and photon emission rate at CEP from effective models as well as from other holographic framework.

\begin{acknowledgments}
We thank Zhibin Li, Hong-An Zeng and Ruixiang Chen for helpful discussions. This work is supported in part by the National Natural Science Foundation of China (NSFC) Grant Nos. 12235016, 12221005, and 12275108, and the Fundamental Research Funds for the Central Universities. 
\end{acknowledgments}

\appendix
\section{Details in the deduction of the photon emission rate and electric conductivity}\label{Appendix}

\subsection{Photon emission rate}
One can integrate over $k$ in \eqref{dGammad3k}, and since $d^3k=k^2dkd\Omega$ we have:
\begin{align*}
    \Gamma&=\int_0^\infty \frac{4\pi k^2dk}{(2\pi)^3}\frac{e^2}{2k}\eta^{\mu\nu}C_{\mu\nu}^{<}=\int \frac{e^2dk}{4\pi^2}k\eta^{\mu\nu}C_{\mu\nu}^{<}
\end{align*}
Taking the derivative to $k$, one can get
\begin{align}
    \frac{d\Gamma}{dk}&=\frac{e^2}{4\pi^2}k\eta^{\mu\nu}C_{\mu\nu}^{<}=\frac{\alpha_{EM}}{\pi}k\eta^{\mu\nu}C_{\mu\nu}^{<}\\
    &=\frac{\alpha_{EM}}{\pi}k\frac{1}{e^{k\beta}-1}\eta^{\mu\nu}\chi_{\mu\nu}=-2\frac{\alpha_{EM}}{\pi}k\frac{1}{e^{k\beta}-1}\eta^{\mu\nu}\mathrm{Im} C_{\mu\nu}^{ret}(K)\\
    &=-4\frac{\alpha_{EM}}{\pi}k\frac{1}{e^{k\beta}-1}\mathrm{Im}\Pi^T\label{emission rate-paa}
\end{align}

\subsection{Electric conductivity}
The simplification from \eqref{C-H's conduct} to \eqref{final expression for conduct} is straightforward:
\begin{align}
     \sigma_Q&=\lim_{\omega\to 0}\frac{e^2}{4T}n_b(k^0){\chi^\mu}_\mu |_{k^0=\omega=|\vec{k}|}\\
     &=-\lim_{\omega\to 0}\frac{e^2}{T}\mathrm{Im}\Pi^T n_b(k^0)|_{k^0=\omega=|\vec{k}|}\\
     &=-\lim_{\omega\to 0}\frac{e^2}{T}\mathrm{Im}\Pi^T \frac{1}{e^{k/T}-1}|_{k^0=\omega=|\vec{k}|}\\
     \Rightarrow \sigma_Q&=-\lim_{\omega\to 0} 4\pi\alpha_{EM}\mathrm{Im}\Pi^T\frac{1}{\omega}
\end{align}

\subsection{Equations of motion and on-shell action}\label{Appendix for eom and on-shell action}

To simpify the deduction, we rewrite the metric \eqref{metric-ansatz} as:
\begin{equation}\label{re-metric}
    ds^2=\frac{1}{\zeta^2(z)}[-g(z)dt^2+\frac{dz^2}{g(z)}+d\vec{x}^2],\quad \frac{L^2e^{2A(z)}}{z^2}\equiv \frac{1}{\zeta^2(z)}
\end{equation}

The equation of motion of the probe action \eqref{probe action} is almost the Maxwell equation, the only difference is the additional coupling $f_Q(\phi)$:
\begin{equation}\label{Maxwell eom}
    \partial_M (\sqrt{-g}f_Q(\phi)F^{MN})=0
\end{equation}

Using \eqref{Maxwell eom}, one can get the equations of motion of every component (with gauge condition $A_z\equiv0$):
\begin{align}
    &\omega(\partial_z A_0)+[g(z)q(\partial_z A_3)]=0\\
    &\partial_z [\frac{f_Q}{\zeta(z)}(\partial_z A_0)]-[\frac{qf_Q}{g(z)\zeta(z)}(q A_0+\omega  A_3)]=0\\
    &\frac{g(z)\zeta(z)}{f_Q}\partial_z [\frac{f_Qg(z)}{\zeta(z)} (\partial_z A_i)]+[\omega^2-q^2 g(z)]A_i=0,\quad i=1,2\label{Transverse}\\ 
    &\partial_z [\frac{f_Qg(z)}{\zeta(z)} (\partial_z A_3)]+\frac{\omega f_Q}{g(z)\zeta(z)}[(q A_0+\omega A_3)]=0
\end{align}
as mentioned above, only the transverse part \eqref{Transverse} is what we really needed (as $\mathrm{Im}\Pi^L$ vanishes for photons). 

In the probe action, the only term that is related to our correlator is:
\begin{equation}
    \int dz d^4x \frac{1}{16\pi G_5}g^{zz}\sqrt{-g}\frac{f_Q(\phi)}{4}\partial_zA_i(z,x)\partial_z A_i(z,x),\quad i=1,2
\end{equation}
after Fourier transformation and integration by parts, we have
\begin{equation}\label{on shell action-1}
    \int d^4k \frac{1}{16\pi G_5}g^{zz}\sqrt{-g}\frac{f_Q(\phi)}{4}A_i(z,-k)\partial_z A_i(z,k)|_{z=0}^{z=z_h}
\end{equation}
where
\begin{align}
    A_i(z,x)&=\int \frac{d^4k}{(2\pi)^4}e^{ik\cdot x}A_i(z,k)\equiv\int \frac{d^4k}{(2\pi)^4}e^{ik\cdot x}A_i(0,k) f_k(z)
\end{align}
and $f_k(0)=1$.

According to \eqref{metric-ansatz}, $g^{zz}\propto g(z)$, on the other hand, $g(z_h)=0$ at $z=z_h$, thus \eqref{on shell action-1} becomes:
\begin{equation}\label{on shell action 2}
    \int d^4k \frac{-1}{16\pi G_5}\frac{f_Q(\phi)}{4}g^{zz}\sqrt{-g}A_i(z,-k)\partial_z A_i(z,k)|_{z=0}
\end{equation}

Using this and the prescription in \cite{DTS-ADS-Prescript}, one can eventually get
\begin{equation}
    \Pi^T=\lim_{z\to 0}\left[ -2\cdot\frac{1}{16\pi G_5}g^{zz}\sqrt{-g}\frac{f_Q(\phi)}{4}\frac{\partial_z A_i(z,k)}{A_i(z,k)} \right]
\end{equation}
where $f_{-k}(z)$ is implicit due to $f_{-k}(0)=1$, and the overall factor "2" here comes from the prescription.

\bibliographystyle{unsrt}
\bibliography{References}

\end{document}